# Angular versus radial correlation effects on momentum distributions of light two-electron ions

Sébastien RAGOT[#], Jean-Michel GILLET[#], Pierre J. BECKER[#@]

[#] *Laboratoire Structure, Propriété et Modélisation des Solides (CNRS, Unité Mixte de Recherche 85-80). École Centrale Paris, Grande Voie des Vignes, 92295 CHATENAY-MALABRY, FRANCE*

[@] *Université Catholique de l'ouest. 1, place André-Leroy. BP808, 49008 ANGERS Cedex 01, FRANCE*

## Abstract

We investigate different correlation mechanisms for two-electron systems and compare their respective effects on various electron distributions. The simplicity of the wave functions used allows for the derivation of closed-form analytical expressions for all electron distributions. Among other features, it is shown that angular and radial correlation mechanisms have opposite effects on Compton profiles at small momenta.

## Introduction

X-ray Compton scattering experiments are a powerful spectroscopy technique for probing the ground-state electronic structure of materials. These experiments lead to the observation of directional Compton profiles, which are closely related to the electron momentum density $n(\mathbf{p})$ [1,2]. The impulse directional profile $J(q,\mathbf{u})$ is obtained from:

$$J(q,\mathbf{u}) = \int n(\mathbf{p}) \delta(q - \mathbf{p}.\mathbf{u}) d\mathbf{p} \qquad (1)$$

where $q$ stands for the momentum variable and $\mathbf{u}$ points along the scattering vector. The current experimental resolution allows to reveal significant deviations to the mean-field Hartree-Fock (HF) or to the local density approximation-based calculations of impulse Compton profiles [3,4]. Beside experimental effects, electron correlation is often invoked to explain these discrepancies. In the case of cubic ionic crystals, the experimental deviations to Hartree-Fock (HF) profiles are mostly isotropic and, interestingly, can show opposite trends, depending on materials studied [3]. We investigated these effects by means of *ab initio* calculations on finite clusters [5] and tried to explain them through simple models [3,5]. We think that the conclusions drawn for isolated ions can be useful for the understanding of the correlation mechanisms in such compounds.

Bound state correlated wave functions have been early investigated for two-electron atoms and molecules, allowing for the prediction of various properties [6,7,8,9]. The correlation effects on Compton profiles of spherical atomic systems:

$$\Delta J(q,\mathbf{u}) \equiv \Delta J(q) = J(q) - J_{HF}(q)$$

have also been investigated [10,11,12], though not as much as position-space properties because of the difficulty inherent to the Fourier-transformation of explicitly correlated wave functions [13]. Such studies have notably shown that the trend of Δ*J* is not systematic. In particular, Meyer and coworkers [14] pointed out that the sign of Δ*J*(*q* = 0) is not always negative, which might involve different correlation mechanisms.

Green and co-workers [15] have, for instance, analyzed the contribution of radial and angular correlation mechanisms to the correlation energy of two-electron ions; their consequences on either the electron pair- or charge densities have also been discussed in the past [16]. For the reason mentioned above, we do know much less regarding momentum space. The purpose of this paper is therefore to draw some qualitative conclusions about the respective role played by of angular and radial correlation mechanisms on Compton profiles of two-electron systems. To this aim, we investigate simple wave functions, which allows for the derivation of analytical expressions for all position or momentum space electron distributions (sect. II). We then illustrate the angular and radial correlation effects on electron pair- (sect. III) and one-electron distributions (sect. IV). Atomic units are used throughout.

## I. Radial versus angular correlation mechanisms

In 1928, Hylleraas [17] determined a very accurate wave function (WF) for the helium atom. The same year, Slater [18] analyzed the correlated hamiltonian $H = h_1 + h_2 + 1/r_{12}$ and concluded that the divergence at the coalescence point ($r_{12} = 0$) yields a cusp in the exact WF, which must in turn exhibit a linear dependence in $r_{12}$ near $r_{12} = 0$. It is well known since Hylleraas's work that including the variable $r_{12}$ in the WF ensures a fast convergence of the correlation energy. Following this idea, a considerable number of accurate analytical WFs have then been proposed (see for example references [15,19,20,21,22,23] and references therein).

Unfortunately, Hylleraas-like wave functions become rapidly prohibitive as the number of electron increases. All the more, the evaluation of the subsequent electron distributions in momentum space clearly makes the use of Hylleraas-like WFs a torment [13]. Nor simple is the resolution of the Schrödinger equation in momentum space [24,25]. The compromise adopted here bypasses these difficulties: it relies on a partial separation of angular and radial effects, as suggested when developing the electronic repulsion term:

$$\frac{1}{r_{12}} = \frac{1}{r_>} + \frac{r_<}{r_>^2}\cos\theta + (1+3\cos2\theta)\frac{r_<^2}{4r_>^3} + ... \tag{2}$$

where $r_>$ denotes the sup of ($r_1, r_2$) and $\theta$ is the inter-electronic angle. The whole potential can thus be developed to first order in $1/r_>$ as $v(r_1, r_2) \equiv v(r_<, r_>) = -(Z-1)/r_> - Z/r_<$. This expression shows that the external part of the electronic cloud experiences a screening of the nucleus due to the internal part, which mechanism invokes radial correlation. In the case of a low nuclear charge ion (like H⁻), this dynamical screening mechanism must be important since *Z* – 1 differs notably from *Z*. The following terms of the development of $v(r_1, r_2)$ further involve *θ*, *i.e.* angular correlation. Notice that angular and radial correlation mechanisms have been extensively investigated, notably through hyperspherical coordinates analyses [26] and the group-theoretical formalism [27].

In order to reflect both radial and angular effects, the trial ground-state WF chosen has the following form:

$$\psi(r_1, r_2) = \sum_{\mu,\nu} c_{\mu\nu} \phi_{1s,\mu}(r_1) \phi_{1s,\nu}(r_2) f_{\mu\nu}(r_1.r_2) \tag{3}$$

where $\phi_{1s,\mu}(r_1)\phi_{1s,\nu}(r_2) = e^{-\mu r_1 - \nu r_2}$. Consider the case $\mu > \nu$: inclusion of the $\phi_{1s,\mu}(r_1)\phi_{1s,\nu}(r_2)$ configuration allows one electron to be close to the nucleus ($\phi_{1s,\mu}$), while the other is farther away ($\phi_{1s,\nu}$). Such a mechanism is often referred to as radial or in-out correlation [28]. The term $f_{\mu\nu}(r_1.r_2)$ has the form $(1 - b_{\mu\nu} r_1.r_2)$, which increases the probability of finding the two electrons on opposite sides of the nucleus [29]. Limiting the expansion of the WF defined in eq. (3) to two different scaling coefficients ($\alpha$ and $\beta$) already involves 8 parameters (or 7 independent parameters). The resulting WF reads:

$$\psi(r_1, r_2) = c_{\alpha\beta}\left(e^{-\alpha r_1 - \beta r_2} + e^{-\beta r_1 - \alpha r_2}\right)(1 - b_{\alpha\beta} r_1.r_2) + c_\alpha e^{-\alpha(r_1+r_2)}(1 - b_\alpha r_1.r_2) + c_\beta e^{-\beta(r_1+r_2)}(1 - b_\beta r_1.r_2) \quad (4)$$

This function will be hereafter denoted by WF (4). Such an approach should obviously be less efficient than Hylleraas's, at least at equal number of parameters. However, as outlined in [30], the lack of odd powers of $r_{12}$ makes possible the analytical calculation of momentum space distributions (see appendix).

We further investigate an angularly correlated wave function of the form:

$$\psi_{ang}(r_1, r_2) = \varphi(r_1)\varphi(r_2)(1 - b\hat{r}_1.\hat{r}_2) \quad (5)$$

$$= \left(\sum_{i=1,3} c_i e^{-\zeta_i r}\right)\left(\sum_{j=1,3} c_j e^{-\zeta_j r}\right)(1 - b\hat{r}_1.\hat{r}_2)$$

where $\hat{r}_1$ and $\hat{r}_2$ are unit vectors. Such a wave function, or WF (5) for short, has however less degrees of freedom than WF (4) for specifically describing angular correlation. Conversely, the WF (4) can be reduced to a function radially correlated only, i.e. WF (6), by setting all $b_k$'s to 0 in eq. (4):

$$\psi_{rad}(r_1, r_2) = c_{\alpha\beta}\left(e^{-\alpha r_1 - \beta r_2} + e^{-\beta r_1 - \alpha r_2}\right) + c_\alpha e^{-\alpha(r_1+r_2)} + c_\beta e^{-\beta(r_1+r_2)} \quad (6)$$

In quantum chemistry, it is customary to define the correlation energy ($\Delta E$) as the exact (non-relativistic, infinite nuclear mass) energy minus the HF one, which is obtained from the best determinantal wave function. In the case of two-electron systems, the spatial part of the ground-state HF wave function $\psi_{HF}$ reduces to a single product of spherical orbitals:

$$\psi_{HF}(r_1, r_2) = \varphi_{HF}(r_1)\varphi_{HF}(r_2) \quad (7)$$

This description obviously ignores any correlated motion of electron positions beyond the mean field approximation. In the following, the HF distributions are calculated from eq. (7), where each orbital $\varphi_{HF}$ are expanded as a sum of three optimally scaled 1s Slater-type orbitals:

$$\varphi(r) = \sum_{i=1,3} c_i e^{-\zeta_i r}$$

The reference HF distributions will hereafter be denoted by the shorthand notation "3$\zeta$–HF".

All investigated wave functions were separately optimized through minimization of $\langle H \rangle$, with $H$ defined as:

$$H = -\tfrac{1}{2}\nabla_1^2 - \tfrac{1}{2}\nabla_2^2 - \left(\frac{Z}{r_1} + \frac{Z}{r_2}\right) + 1/r_{12} \quad (8)$$

The various resulting correlation energies are reported in table 1. First, we observe that the radial correlation mechanism is dominant in the case of H⁻. Notice that the amount of radial correlation energy obtained from WF (6) closely follows that of Green and co-workers [15], while WF (5) clearly underestimates the angular contribution. The WF (4) turns out to be more appropriate for H⁻, for which the mutual screening of electrons is proportionally more important than for higher Z ions. Comparing our results to those obtained with Hylleraas-like WFs indicates that the accuracy of WF (4) is superior to two-parameter Hylleraas wave functions, but is inferior to three-parameter ones [31], at least for $Z = 1$ to 3.

Table 2 lists the optimal parameters for WF (4), calculated for H$^-$ to Li$^+$. The evolution of the parameters with Z clearly illustrates the fade of radial correlation on the wave function itself. First, the Slater exponents ($\alpha,\beta$) increase with Z but the ratio of their difference to their average decreases. Furthermore, the weight of in-out configurations ($c_{\alpha\beta}$) continuously decreases while the $b_k$'s increase ($k = \alpha, \beta$ or $\alpha\beta$). Notice that negative $b_\beta$'s are associated with negative $c_\beta$'s.

## II. Electron pair distributions

Electron pair densities of atomic systems (and various related distributions) have been a subject of constant research during the last decade (in both position and momentum spaces, see for instance ref. [32]). The ground-state electron pair density of a two-electron system is defined as

$$P(\mathbf{r}_1,\mathbf{r}_2) = |\psi(\mathbf{r}_1,\mathbf{r}_2)|^2 \tag{9}$$

In other words, $P(\mathbf{r}_1,\mathbf{r}_2)$ is the probability density of finding electron 1 at position $\mathbf{r}_1$ while electron 2 is located at position $\mathbf{r}_2$, and fulfills the normalization condition:

$$\int P(\mathbf{r}_1,\mathbf{r}_2) d\mathbf{r}_1 d\mathbf{r}_2 = 1 \tag{10}$$

Note that in a correlated description, $P(\mathbf{r}_1,\mathbf{r}_2)$ depends explicitly on $r_1$, $r_2$ and the inter-electronic angle coordinate $\theta$, so that it can be denoted by $P(\mathbf{r}_1,\mathbf{r}_2) \equiv P(r_1,r_2,\theta)$. Given the symmetry of the system, we can replace the 2-particle volume element $d\mathbf{r}_1 d\mathbf{r}_2$ in (10) by $(4\pi r_1^2)(2\pi r_2^2)\sin\theta\, d\theta\, dr_1\, dr_2$. This led us to consider the two functions $D(r_2,\theta)$ and $A(\theta)$, respectively defined as:

$$D(r_2,\theta) = \int P(r_1,r_2,\theta) 8\pi^2 r_1^2 r_2^2 \sin\theta\, dr_1 \tag{11}$$

and

$$A(\theta) = \int P(r_1,r_2,\theta) 8\pi^2 r_1^2 r_2^2\, dr_1 dr_2 \tag{12}$$

These functions satisfy the following normalization conditions:

$$\int D(r_2,\theta) d\theta dr_2 = 1$$

and

$$\int A(\theta) \sin\theta\, d\theta = 1$$

The function $A(\theta)$ is the probability density of finding a pair with an inter-electronic angle $\theta$. Note that in the HF description, $P_{HF}$ does not depend on $\theta$, from (7) and (9), and so $A_{HF}(\theta)$ reduces to

$$A_{HF}(\theta) = \frac{1}{2} \tag{13}$$

The function $D(r_2,\theta)$ might be seen as the $\theta$-dependent distribution of the distance of electron 2 to the nucleus. The particular case $\theta = 0$ shall give us the opportunity to illustrate the radial correlation phenomenon. In fig. 1, we compare the $D(r_2,\theta=0)$ distributions for H$^-$, He and Li$^+$, as obtained from WF (4). The case of H$^-$ exhibits one sharp and one broad peak, a fingerprint of the dynamical screening of the nucleus evoked in section II. As expected, the dynamical screening mechanism fades for Higher Z ions: no distinct peaks can be seen for He and Li$^+$ at the scale of the plot. Figure 2 focuses on the $D(r_2, \theta = 0)$ distribution for H$^-$. The sharp and the broad peaks are for instance well separated by a vertical axis located at the position of the average distance of an electron to the nucleus. Roughly speaking, the broad peak reflects the loosely bound residual distribution of electron 2 which "sees" a nucleus screened by electron 1. The 3ζ–HF distribution is also displayed, for comparison. It takes higher values due to the fact that it neglects angular correlation too [ $A_{HF}(\theta) = \frac{1}{2}$ ], which mechanism lowers the pair density near $\theta = 0$.

Figure 3 shows a polar plot of the correlated angular probability densities $A(\theta)$, obtained from WF (4). They all have a maximum centered on $\theta = \pi$, which feature is consistent with the intuitive classical picture (angular correlation must at least partially reject electrons on opposite sides of the nucleus [33]). The phenomenon is particularly accentuated in the case of H$^-$, the angular density of which shows a dip at $\theta = 0$.

## III. Comparison of one-electron distributions

As we shall see now, the magnitude of the correlation contribution to one-electron distributions drastically decreases as the nucleus charge increases. The one-electron density matrix is defined as:

$$\rho_1(\mathbf{r}_1, \mathbf{r}_1') = 2\int \psi(\mathbf{r}_1, \mathbf{r}_2)\psi^*(\mathbf{r}_1', \mathbf{r}_2) d\mathbf{r}_2 \qquad (14)$$

The electron density $\rho(\mathbf{r})$ can be equivalently obtained by taking the diagonal element of $\rho_1(\mathbf{r}, \mathbf{r}')$, so that $\rho(\mathbf{r}) = \rho_1(\mathbf{r}, \mathbf{r})$, or by integrating $P(\mathbf{r}_1, \mathbf{r}_2)$ over $\mathbf{r}_2$ (and multiplying it by 2). Given the spherical symmetry, we further have $\rho(\mathbf{r}) = \rho(r)$. The electron distribution in position space is thus:

$$D(r) = 4\pi r^2 \rho(r) \qquad (15)$$

It is related to eq. (11) through $D(r) = 2\int_0^\pi D(r, \theta) d\theta$. The correlation contribution to the position-space distribution expresses as:

$$\Delta D(r) = 4\pi r^2 [\rho(r) - \rho_{HF}(r)] \qquad (16)$$

The calculation of the momentum density $n(\mathbf{p})$ requires knowledge of $\rho_1(\mathbf{r}, \mathbf{r}')$ for $\mathbf{r} \neq \mathbf{r}'$:

$$n(\mathbf{p}) = \frac{1}{(2\pi)^3} \int \rho_1(\mathbf{r}, \mathbf{r}') e^{i\mathbf{p}.(\mathbf{r}-\mathbf{r}')} d\mathbf{r} d\mathbf{r}' \qquad (17)$$

One can further define the momentum distribution: $I(p) = 4\pi p^2 n(p)$. The impulse Compton profile $J(q)$, defined by eq. (1), can be rewritten as:

$$J(q) = 2\pi \int_{|q|}^\infty n(p) p\, dp = \frac{1}{2}\int_{|q|}^\infty I(p)/p\, dp \qquad (18)$$

The correlated distributions have been computed from the optimal WFs (4), (5) and (6), while the reference HF distributions derive from (7). We have checked that the correlation contributions compare well with more sophisticated ones in the case of the H$^-$. in particular, the correlation Compton profile of H$^-$ was found to be very close to that obtained from a Gaussian94 [34] calculation with an extended basis-set (see fig. 4). The agreement with the benchmark correlation profile of Regiert and Thakkar [35] is even better (this last has been obtained from a wave function expanded in 30 optimized gaussian geminals). For comparison, we have included various one-electron momentum properties for H$^-$ and He in table 3, which confirms that the WF (4) is better suited for H$^-$ than for He. In particular, The WF (4) overestimates the value $n(p = 0)$ in the case of He.

We also compared radial, angular and global correlation effects on one-electron distributions for H- and He. Correlation contributions are illustrated in fig. 5 for position space distributions (left column) and Compton profiles (right column). Larger magnitudes are associated to a lower nuclear charge ion. This suggests that, in more complex systems, the most visible correlation-induced deformations shall be associated with valence electrons. We further plotted the difference between the profiles derived respectively from WFs (4) and (6). The resulting deformations can however not be ascribed to the only angular effects, since correlation is not fully separable in purely angular and radial correlation [15]. They are nevertheless similar to these computed from WF (5), regardless of the magnitudes. Radial correlation spreads out electron distributions in both position and momentum spaces. This causes the radial correlation contributions to be positive at small and large values of $r$ or $q$. It follows that radial correlation alone overestimates the Compton profile at low momenta (this holds for higher $Z$ ions). Conversely, the only angular correlation allows the electron cloud to get closer to the nucleus in average, brings a negative contribution to $J(0)$ and slightly shifts the Compton profile towards higher momenta. Thus, angular and radial correlation mechanisms have opposite effects on $J(0)$ but the deformations in momentum space are in all cases positive at large $q$ [35, 36]. In other words, the mutual correlation of electrons makes them move faster in average (see also the $\langle p^n \rangle$ values, $n \geq 1$, reported in table 3).

Besides, the reported deformations reflect the error associated with the mean field approximation: for instance, one can conclude that, when radial correlation dominates, the HF approximation overestimates electron distributions near $\langle x \rangle$, where $x = r, q$. This interpretation holds for the momentum density of light two-electrons systems but is not straightforwardly extensible to heavier ions. We finally point out that the correlation profile of He obtained from WF (4) is positive at low $q$, in contradiction with the results of Regiert and Thakkar [35]. This shortcoming is due to the low angular expansion of WF (4), which underestimates the subsequent angular contribution to the correlation energy (see table 1). In facts, it was shown in [15] that the angular correlation slightly dominates in the case of He (table 1), so that the global correlation contribution to $J(0)$ should be negative at small momenta.

## IV. Conclusion

Simple wave functions allowed us for illustrating the effects of different correlation mechanisms on electron distributions of some two-electron systems. Although not very sophisticated, the functions chosen permit recover up to 93% of the exact correlation energies. They can further be analytically formulated in closed-form in both position and momentum spaces. Correlation is shown to bring important deformation magnitudes on the electron distributions of H-; these magnitudes, however, drastically decrease for higher Z ions. Besides, the dynamical screening mechanism has important consequences on the structure of the electron pair density of the anion. For all studied species, the radial correlation mechanism was found to widen the one-electron distributions in both position and momentum space. Conversely, the only angular correlation mechanism slightly shifts the momentum distributions towards high momenta and allows the electron cloud to get closer to the nucleus. The correlation contributions to the Compton profile are in all cases positive at large momenta, which reflects the fact that correlation makes electrons move faster in average [36]. The angular correlation mechanism is shown to bring a negative contribution to $\Delta J(0)$. Such a conclusion is consistent with the observed trend for the correlation profiles of the Be-isoelectronic sequence [11], where the near degeneracy between the $2s$ and $2p$ states enhances the angular correlation in the valence shell. Note finally that correlation could be significant in ionic crystals [3] (like LiH and MgO), where anionic electrons are mostly confined to a finite region of space, as compared with free anions. The competition between radial and angular correlation mechanism may thus be responsible for the observed experimental trends, that is, $\Delta J(0) < 0$ for LiH and $\Delta J(0) > 0$ for MgO.

# Appendix

Consider the position space wave function (3):

$$\psi(r_1, r_2) = \sum_{\mu,\nu} c_{\mu\nu} \phi_{1s,\mu}(r_1) \phi_{1s,\nu}(r_2) f_{\mu\nu}(r_1, r_2)$$

of which momentum space counterpart is obtained from the double Fourier-transform:

$$\tilde{\psi}(p_1, p_2) = \frac{1}{(2\pi)^3} \sum_{\mu,\nu} c_{\mu\nu} \int f_{\mu\nu}(r_1, r_2) \phi_{1s,\mu}(r_1) \phi_{1s,\nu}(r_2) e^{-i(p_1 \cdot r_1 + p_2 \cdot r_2)} dr_1 dr_2$$

replacing $f_{\mu\nu}(r_1, r_2) = (1 - b_{\mu\nu} r_1 \cdot r_2)$ by $\hat{f}_{\mu\nu}(p_1, p_2) = 1 + b_{\mu\nu} \nabla_{p_1} \cdot \nabla_{p_2}$ yields:

$$\tilde{\psi}(p_1, p_2) = \sum_{\mu,\nu} c_{\mu\nu} \hat{f}_{\mu\nu}(p_1, p_2) \tilde{\phi}_{1s,\mu}(p_1) \tilde{\phi}_{1s,\nu}(p_2)$$

where $\tilde{\phi}_{1s,\mu}$ is the Fourier-transform of the position space orbital $\phi_{1s,\mu}$. Finally, the momentum space wave function writes explicitly as:

$$\tilde{\psi}_{\mu\nu}(p_1, p_2) = \sum_{\mu,\nu} c_{\mu\nu} \left(1 + \frac{16 b_{\mu\nu} p_1 \cdot p_2}{(p_1^2 + \mu^2)(p_2^2 + \nu^2)}\right) \left[\frac{8\mu\nu}{\pi(p_1^2 + \mu^2)^2 (p_2^2 + \nu^2)^2}\right]$$

The (mainly angular) correlation factor thus increases the probability of finding electrons with collinear momenta. The one-electron density matrix is defined as:

$$\rho_1(r_1, r_1') = 2 \int \psi(r_1, r_2) \psi^*(r_1', r_2) dr_2$$
$$= 2 \sum_{\alpha,\beta,\delta,\gamma} c_{\alpha\beta} c_{\delta\gamma} \phi_{1s,\alpha}(r_1) \phi_{1s,\delta}(r_1') \int \phi_{1s,\beta}(r_2) \phi_{1s,\gamma}(r_2') f_{\alpha\beta}(r_1, r_2) f_{\delta\gamma}(r_1', r_2') dr_2$$

Or in compact form: $\rho_1(r_1, r_1') = 2 \sum_{\alpha,\beta,\delta,\gamma} c_{\alpha\beta} c_{\delta\gamma} \rho_{1,\alpha\beta\delta\gamma}(r_1, r_1')$. We find explicitly:

$$\rho_1(r_1, r_1') = 2 \sum_{\alpha,\beta,\delta,\gamma} c_{\alpha\beta} c_{\delta\gamma} \left\{ \frac{8(\alpha\beta\delta\gamma)^{3/2} \left[4 b_{\alpha\beta} b_{\delta\gamma} r_1 \cdot r_1' + (\beta+\gamma)^2\right] e^{-(\alpha r_1 + \delta r_1')}}{\pi (\beta+\gamma)^5} \right\}$$

and finally:

$$n(p) = 2 \sum_{\alpha,\beta,\delta,\gamma} c_{\alpha\beta} c_{\delta\gamma} n_{\alpha\beta\delta\gamma}(p)$$
$$= 2 \sum_{\alpha,\beta,\delta,\gamma} c_{\alpha\beta} c_{\delta\gamma} \left\{ \frac{64 \alpha\beta (\alpha\beta\delta\gamma)^{3/2} \left[64 b_{\alpha\beta} b_{\delta\gamma} p^2 + (\beta+\gamma)^2 (p^2 + \alpha^2)(p^2 + \delta^2)\right]}{\pi^2 (\beta+\gamma)^5 (p^2 + \alpha^2)^3 (p^2 + \delta^2)^3} \right\}$$

TABLE CAPTIONS & TABLE

Table 1: Component analysis of correlation energies of some two-electron ions (Z = 1, 3). Comparison between the results obtained from WFs (4), (5) and (6) and these of Green and co-workers [15]. Results of Green and co-workers also contain a mixed contribution which involves both angular and radial correlation between electron positions. The reference correlated and HF energies have been taken from ref. 37 and 38, respectively. The reference correlation energy "$\Delta E_{Exact}$" is thus $E$[ref. 37] - $E_{HF}$[ref. 38].

Table 2: Optimal parameters obtained for the (normalized) wave functions (4).

Table 3: Comparison of one-electron momentum properties of H⁻ and He. Note that the current accuracy of large-scale variational calculations of ground-state energies of two-electron ions is superior to 10 digits.

Table 1

|  | $H^-$ | He | $Li^+$ |
|---|---|---|---|
| *$\Delta E$ components from this work:* | | | |
| $\Delta E_{Rad}/\Delta E_{Exact}$ = $(E_{Rad}[WF(6)]-E_{HF})/\Delta E_{Exact}$ | 0.639 | 0.361 | 0.313 |
| $\Delta E_{Ang}/\Delta E_{Exact}$ = $(E_{Ang}[WF(5)]-E_{HF})/\Delta E_{Exact}$ | 0.196 | 0.239 | 0.241 |
| $\Delta E_{Total}$ = $(E_{Total}[WF(4)]-E_{HF})/\Delta E_{Exact}$ | 0.926 | 0.841 | 0.800 |
| *$\Delta E$ components from ref. [15]:* | | | |
| $\Delta E_{Rad}/\Delta E_{Exact}$ = $(E_{Rad}[ref. 15]-E_{HF})/\Delta E_{Exact}$ | 0.654 | 0.418 | 0.336 |
| $\Delta E_{Ang}/\Delta E_{Exact}$ = $(E_{Ang}[ref. 15]-E_{HF})/\Delta E_{Exact}$ | 0.385 | 0.541 | 0.601 |
| $\Delta E_{Mixed}/\Delta E_{Exact}$ = $(E_{Mixed}[ref. 15]-E_{HF})/\Delta E_{Exact}$ | -0.085 | -0.014 | -0.096 |
| $\Delta E_{Total}/\Delta E_{Exact}$ = $(E_{Total}[ref. 15]-E_{HF})/\Delta E_{Exact}$ | 0.954 | 0.945 | 0.927 |

Table 2

|  | $\alpha$ | $\beta$ | $c_\alpha$ | $c_\beta$ | $c_{\alpha\beta}$ | $b_\alpha$ | $b_\beta$ | $b_{\alpha\beta}$ |
|---|---|---|---|---|---|---|---|---|
| $H^-$ | 0.361721 | 1.068192 | 0.0343866 | -0.0845432 | 0.609047 | 0.039944 | -0.774243 | 0.001635 |
| He | 1.350225 | 2.459764 | 0.305753 | -0.13461 | 0.440822 | 0.112874 | -1.956528 | -0.08828 |
| $Li^+$ | 2.39972 | 4.19635 | 0.576037 | -0.124237 | 0.288935 | 0.144556 | -3.43266 | -0.231353 |

Table 3

|  | $E$ | $T = \langle p^2 \rangle/2$ | $n(0)$ | $\langle p^{-2} \rangle$ | $\langle p^{-1} \rangle = 2J(0)$ | $\langle p \rangle$ | $\langle p^2 \rangle$ | $\langle p^3 \rangle$ |
|---|---|---|---|---|---|---|---|---|
| $H^-$ | | | | | | | | |
| $3\zeta$–HF | -0.487818 | 0.488045 | 11.7745 | 34.5709 | 5.99912 | 1.09812 | 0.97609 | 1.45783 |
| WF (4) | -0.524797 | 0.52408 | 16.9548 | 42.3885 | 6.43069 | 1.11465 | 1.05239 | 1.63137 |
| 30 GG[a] | -0.527698 | 0.52770 | 17.4210 | 42.90 | 6.4456 | 1.1147 | 1.0554 | 1.6580 |
| Pekeris[b] | -0.527751 | 0.527751 | - | - | - | - | 1.0555 | - |
| He | | | | | | | | |
| $3\zeta$–HF | -2.86157 | 2.86157 | 0.43720 | 4.08278 | 2.13984 | 2.79897 | 5.72315 | 17.9833 |
| WF (4) | -2.89705 | 2.89561 | 0.46006 | 4.18064 | 2.15328 | 2.80927 | 5.79122 | 18.2476 |
| 40 GG[a] | -2.903701 | 2.90370 | 0.44273 | 4.0986 | 2.1386 | 2.814 59 | 5.8074 | 18.4056 |
| Pekeris[b] | -2.903724 | 2.90372 | - | - | - | - | 5.80744 | - |

[a] GG stands for gaussian geminals, from the work of Regiert and Thakkar [35]
[b] From the accurate wave functions of Pekeris. H-: see ref. [8]. He: ref. [7].

FIGURE CAPTIONS & FIGURES

Figure 1: $D(r_2, \theta = 0)$ distributions [function (11) of text, computed from (4)] for H⁻, He and Li⁺. *Full line*: H⁻. *Dashed*: He. *Large dashed*: Li⁺.

Figure 2: $D(r_2, \theta = 0)$ distributions [(function (11) of text] for H⁻. *Full line*: correlated distribution computed from WF (4). *Dotted-large dashed*: 3ζ−HF distribution, computed from (7). *Vertical dotted line*: position of $\langle r_1 \rangle$.

Figure 3: Polar plots of angular probability densities [function (12) of text] for H⁻, He and Li⁺. *Full line*: H⁻. *Dashed*: He. *Large dashed*: Li⁺. *Dotted-large dashed*: HF functions (identical for H⁻, He and Li⁺).

Figure 4: Correlation contribution to the Compton profile of H⁻. *Full line*: computed from WF (4). *Dashed*: computed from Gaussian94 outputs (CI calculation within AUG-cc-pV5Z basis-set [39]). *Dotted*: from the gaussian geminal wave function of Regiert and Thakkar [35]. The reference HF profiles are computed from WF (7).

Figure 5: Correlation effects on one-electron distributions for H⁻ (1ʳˢᵗ raw) and He (2ⁿᵈ raw). *Left*: correlation contributions to position space distributions (function (16) of text). *Right*: correlation contributions to Compton profiles. *Full line*: from WF (4). *Dotted-dashed*: from angularly correlated WF (5) *Dashed*: from radialy correlated WF (6). *Dotted*: difference between global and radial correlation profile. *Vertical dashed line*: position of $\langle x \rangle$, $x = r, q$. The reference HF distributions are computed from WF (7).

Figure 1

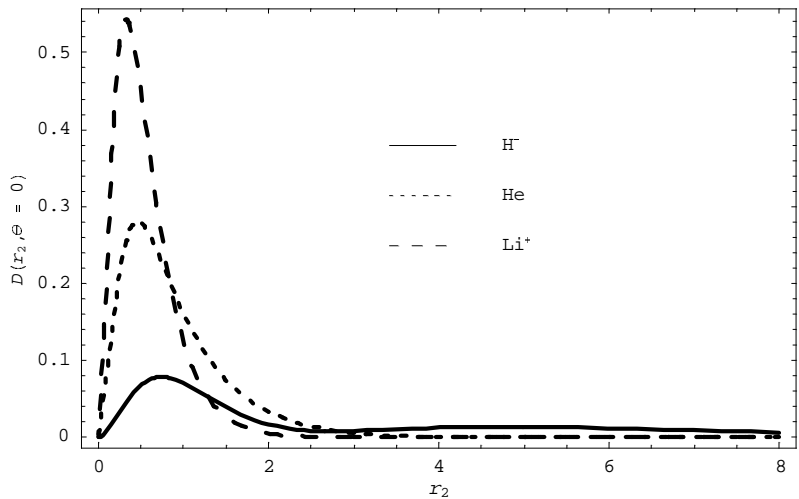

Figure 2

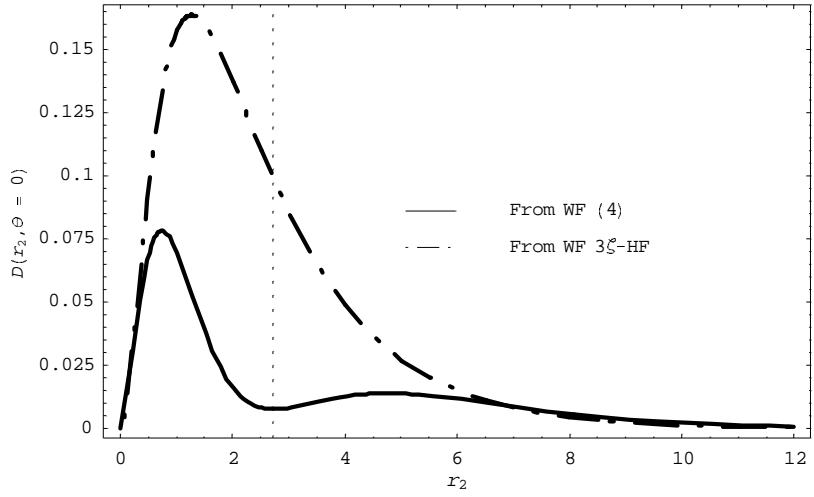

Figure 3

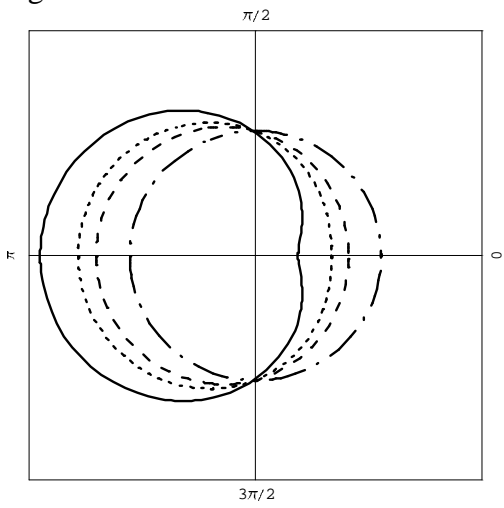

Figure 4

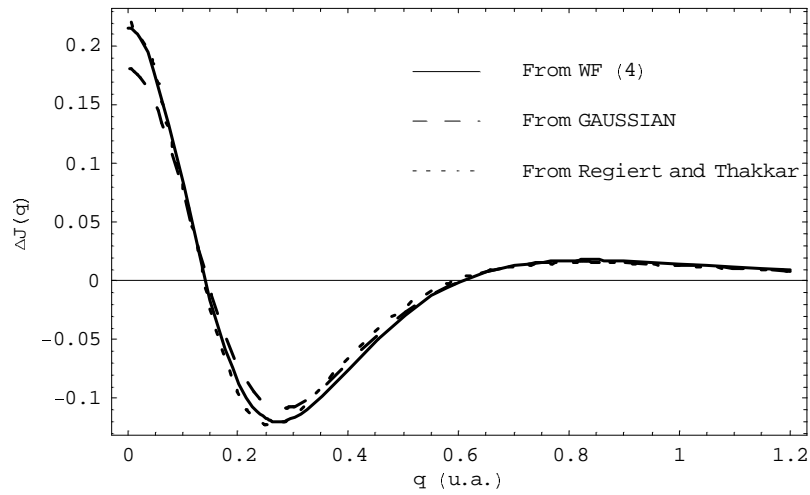

Figure 5

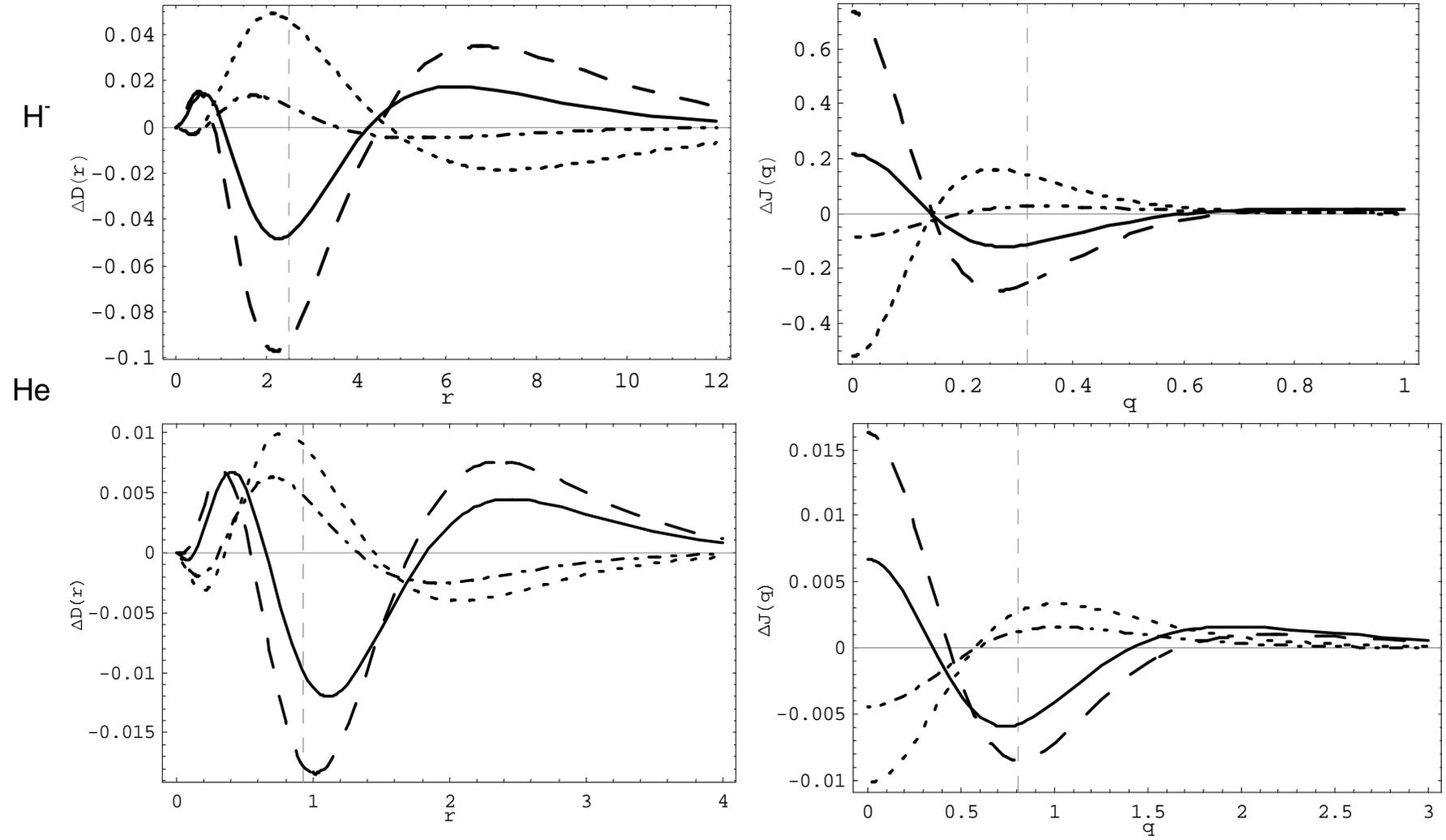

H⁻

He